\documentclass[a4paper,12pt]{article}
\usepackage[english]{babel}
\usepackage{amsfonts,amssymb,amsmath}

\begin{document}

 \centerline{\bf\Large Formation of compression waves} 
\vskip 1mm 
 \centerline{\bf\Large with multiscale asymptotics}
\vskip 1mm 
 \centerline{\bf\Large in the Burgers and KdV models}

 \vskip 5mm 
 \centerline{\bf S.V.~Zakharov} 

 \vskip 5mm 

\begin{center}
Institute of Mathematics and Mechanics,\\
Ural Branch of the Russian Academy of Sciences,\\
16, S.Kovalevskaja street, 620990, Ekaterinburg, Russia
\end{center}
 
\vskip 5mm 

\textbf{Abstract.}
The Cauchy problem for the Burgers equation
with a small dissipation and an initial weak discontinuity
and the Cauchy problem with a large initial gradient
for a quasilinear parabolic equation 
and for the Korteweg--de Vries (KdV) equation are considered.
Multiscale asymptotics of solutions corresponding
to shock waves are constructed.
Some results can also be applied to rarefaction waves.


 
\vskip 5mm 

\section{Wave with an initial weak discontinuity}

A simplest model of the motion of continuum,
which takes into account nonlinear effects and dissipation,
 is the equation of nonlinear diffusion 
\begin{equation}\label{eqw}  
 \frac{\partial u}{\partial t} + 
 u\frac{\partial u}{\partial x} = 
 \varepsilon \frac{\partial^2 u}{\partial x^2},
\qquad \varepsilon>0,
\end{equation}
for the first time presented by J.~Burgers~\cite{bu}. 
 This equation  is used in studying
 the evolution of a wide class of physical systems and probabilistic
processes, for example, acoustic waves in fluid and gas~\cite{Gbw,ib}. 
The problem of shock wave formation 
(for equation~(\ref{eqw}) with $\varepsilon>0$)
is investigated in~\cite{ib,vm4,sp}.
A perturbed weak discontinuity is considered in~\cite{BKdV}.
The problem of shock wave propagation for the Hopf type
equation ($\varepsilon=0$) is studied in~\cite{DM1,DSh}.

Below, we briefly state results of paper~\cite{vm4}
for the following initial data:
	\begin{equation}\label{ivp}
	u(x,-1,\varepsilon) = - (x + ax^2) \Theta(-x),\quad x\in\mathbb{R}^1,
	\end{equation}
where $a>0$, $\Theta(x)\vert_{x\geqslant 0}=1$, $\Theta(x)\vert_{x<0}=0$.
The solution of the limit ($\varepsilon=0$) equation 
is found using the method of characteristics: 
 	$$ 
 	u_0(x,t) =\left\{ 
 \begin{array}{ll} 
 \displaystyle\frac{ 2a (1 + t) x +
t + \sqrt{t^2 -4a (1 + t) x}} 
 	{2a (1 + t)^2}, & x < s(t)\Theta(t), \\ 
 	0, & x > s(t)\Theta(t).\\ 
 	 \end{array}\right. 
 	$$ 
The function $u_0$ has a jump discontinuity on
the line $x=s(t)$, 
 where 	\begin{equation}\label{swc} 
 	 s(t) \equiv \frac{(3t + 4)^{3/2} - 9t - 8}{9a(1+t)}, 
 \qquad 
 s(t) = \frac{3 t^2}{16 a} - \frac{27 t^3}{128 a} + O(t^4), 
 \quad t\to 0. 
 \end{equation} 
The solution of problem~(\ref{eqw})--(\ref{ivp}) is
given by the following expression: 
 	\begin{equation}\label{ipsi} 
 	u(x,t,\varepsilon) = -2\varepsilon 
 	[\Psi(x,t,\varepsilon)]^{-1}\Psi_x(x,t,\varepsilon), 
 \end{equation} 
 where 
 	$$ 
 	\Psi(x,t,\varepsilon) = \frac{1}{2\sqrt{\pi (1 + t)}} 
 	\int\limits_{-\infty}^{0} \exp 
 	\left\{ \frac{1}{\varepsilon} \left[ 
 	-\frac{(x - y)^2}{4\,(1 + t)} + \frac{y^2}{4} 
 	+ \frac{ay^3}{6} \right]\right\} \, dy ~+ 
 	\phantom{=====} 
 	$$ 
 	$$ 
 	\phantom{=========} 
 	+~\frac{1}{2\sqrt{\pi (1 + t)}} 
 	\int\limits_{0}^{\infty}\exp 
 \left\{ \frac{1}{\varepsilon} \left[ 
 	-\frac{(x - y)^2}{4\,(1 + t)} \right]\right\} \,
dy. 
 	$$ 
 Taking into account these formulas, we write
expression~(\ref{ipsi}) in the form
 	\begin{equation}\label{ubypsi} 
 	u(x,t,\varepsilon) = 
 - \frac{\Psi^0(x,t,\varepsilon)} 
 {\phantom{\Big[}\Psi^-(x,t,\varepsilon) 
 + \Psi^+(x,t,\varepsilon)\phantom{\Big]}}, 
 \end{equation} 
 where $$ 
 \Psi^-(x,t,\varepsilon) = 
 \int\limits_{-\infty}^{0} 
 \exp \left\{ \frac{F^{-}(y,x,t)}{\varepsilon}\right\}
dy, 
 \qquad 
 F^{-}(y,x,t) = - \frac{(y - x)^2}{4 (1 + t)} + 
 \frac{y^2}{4} + \frac{a y^3}{6}, 
 $$ 
 	\begin{equation}\label{psip} 
 	\Psi^+(x,t,\varepsilon)=2\sqrt{\varepsilon\pi(1+t)} 
 	- \frac{2\varepsilon(1+t)}{x} 
 	\exp\left\{ -\frac{\zeta^2}{4(1+t)}\right\} 
 	+ \sqrt{\varepsilon(1+t)} 
 	\int\limits_{-\infty}^{-\frac{\zeta}{2\sqrt{1+t}}} 
 	\frac{e^{-z^2}}{z^2}dz,
 \end{equation} 
$\zeta = x / \sqrt{\varepsilon}$,
 	$$ 
 	\Psi^0(x,t,\varepsilon) = \int\limits_{-\infty}^{0} (y+ay^2) 
 \exp \left\{ \frac{F^{-}(y,x,t)}{\varepsilon}\right\} dy. 
 $$ 
 
The integral $\Psi^-(x,t,\varepsilon)$ 
can be written in the form of the sum
 	$$ 
 	\Psi^-(x,t,\varepsilon) = 
 \Psi^-_s(x,t,\varepsilon) + \Psi^-_b(x,t,\varepsilon) 
 = \int\limits_{-\infty}^{y^{+}(x,t)}e^{F^-/\varepsilon} dy 
 + \int\limits_{y^+(x,t)}^{0}e^{F^-/\varepsilon} dy, 
 $$ 
 where $y^+(x,t)$ is the point of a local minimum
of $F^-(y,x,t)$  in the variable~$y$. 
The asymptotics of $\Psi^-_s(x,t,\varepsilon)$ 
 is found by Laplace's method. 
 Therefore,  we need expressions 
 	\begin{equation}\label{sps} 
 	y^\mp(x,t) = - \frac{t \pm \sqrt{t^2 - 4ax(1 + t)}}{2a (1 + t)}
 	\end{equation}	 
for extremal
points of the function $F^-$ in the variable~$y$ 
 (for fixed~$x$ and~$t$) 
 and values of the function $F^-$ and its derivatives at
these points. 
 
 Substituting $y = y^{\mp}(x,t)$ into expressions
for $F_{yy}^-(y,x,t)$ and $F^-(y,x,t)$, we
have
\begin{equation}\label{fy2} 
 F^-_{yy}(y^\mp(x,t),x,t) = \mp 
 \frac{R(x,t)}{2 (1 + t)}, 
 	\quad \mbox{where}\quad 
 R(x,t) = \sqrt{t^2 - 4ax (1 + t)}, 
 \end{equation} 
 	\begin{equation}\label{fmy} 
 	F^-(y^\mp(x,t),x,t) = - \frac{xt + 3ax^2(1 +
t) - \displaystyle\frac{t \pm R(x,t)}{2a(1 +
t)}[R(x,t)]^2}{12 a (1 + t)^2}. 
 \end{equation} 
Thus, 
 	\begin{equation}\label{psimsa} 
 	\Psi^-_s(x,t,\varepsilon) = 
 \sqrt{2\pi\varepsilon} 
 \exp\left[ \frac{F^-(y^-(x,t),x,t)}{\varepsilon}\right] 
 \sum\limits_{j=0}^{\infty} 
 \varepsilon^{j} \gamma_{j} a^{2j} [H(x,t)]^{6j+1}, 
 \end{equation} 
 where 
$$ 
 	H(x,t) = \frac{1}{\sqrt{-F^{-}_{yy}(y^-(x,t),x,t)}}, 
 \qquad 
 \gamma_{0} = 1, \quad \gamma_{1} = 5/24. 
 $$ 

 Let us introduce a stretched variable $\sigma = \displaystyle\frac{x - s(t)}{\varepsilon}$, 
 where $s(t)$ is from~(\ref{swc}), 
 We make the change $x=s(t)+\varepsilon\sigma$ 
 in expressions for $R(x,t)$, $H(x,t)$,
and $y^{\pm}(x,t)$: 
 	\begin{equation}\label{rxt} 
 	R(x,t)\Big\vert_{x=s(t)+\varepsilon\sigma} 
 = \frac{[r(t)]^2}{3} + 
 \sum\limits_{m=1}^{N-1}(\sigma\varepsilon)^{m}
r_m(t)  + O(\varepsilon^{\alpha N}), 
 \quad r_m(t) = O(t^{1-2m}), 
 \end{equation} 
 	\begin{equation}\label{ha} 
 	H(x,t)\Big\vert_{x=s(t)+\varepsilon\sigma} 
 = \sqrt{\frac{2(1+t)}{R(s(t)+\varepsilon\sigma,t)}} 
 = \sqrt{6(1+t)} \sum\limits_{m = 0}^{N-1}
c_m 
 \frac{[a (1 + t) \sigma \varepsilon]^m}{[r(t)]^{4m + 1}} 
 + O\left( \varepsilon^{\alpha N} \right), 
 \end{equation} 
 where $r(t) = \sqrt{3t + 4 - 2\sqrt{3t + 4}}$, $c_0 = 1$, $c_1 = 9$; 
 	\begin{equation}\label{ypms} 
 	y^{\pm}(x,t)\Big\vert_{x=s(t)+\varepsilon\sigma} 
 = \sum\limits_{m=0}^{N-1} (\sigma\varepsilon)^m 
 Y^{\pm}_{m}(t) + O(\varepsilon^{\alpha N}), 
 \quad 
 Y^{\pm}_{m}(t) = O(t^{1-2m}). 
 \end{equation} 
Using expansion~(\ref{rxt}),
 let us pass to variable $\sigma$ in
formula~(\ref{fmy}) 
 	\begin{equation}\label{fma} 
 	F^-(y^-(x,t),x,t)\Big\vert_{x=s(t)+\varepsilon\sigma} 
 	= - \mu(t)\sigma\varepsilon + 
 	\sum\limits_{m=2}^{N-1} (\varepsilon\sigma)^m F^-_{-,m}(t) 
 	+ O(t^3 \varepsilon^{\alpha N}), 
 	\end{equation} 
 	\begin{equation}\label{fpa} 
 	F^-(y^+(x,t),x,t)\Big\vert_{x=s(t)+\varepsilon\sigma} 
 = - \frac{1}{324 a^2}\frac{[r(t)]^6}{(1 + t)^3} 
 + \sum\limits_{m=1}^{N-1} (\varepsilon\sigma)^m F^-_{+,m}(t) 
 	+ O(t^3 \varepsilon^{\alpha N}),	 
 \end{equation} 
 	$$ 
 	\quad\mbox{where}\quad 
 	F^-_{\pm,m}(t) = O(t^{3-2m}), 
 \qquad 
 \mu(t) = s'(t) 
 = \frac{(3t + 1)\sqrt{3t + 4} - 2}{18 a (1 + t)^2}. 
 	$$ 
 Using expansions~(\ref{psimsa}) and~(\ref{ha}), 
 we obtain 	$$ 
 	\Psi^-_s(x,t,\varepsilon) = 
 \sqrt{2\pi\varepsilon} 
 \exp\left[ \frac{F^-(y^-(x,t),x,t)}{\varepsilon} \right] 
 \frac{\sqrt{6(1+t)}}{r(t)} 
 \left\{ 
 1 + \sum\limits_{m=1}^{N-1} \varepsilon^m \Psi^{-}_{s,m}(\sigma,t) 
 + O(\varepsilon^{\alpha N}) 
 \right\}, 
 $$ 
 where $\Psi^{-}_{s,m}\in\Pi_m$. 
 To describe the behavior of coefficients of some
expansion, we used the class of functions 	
$$ 
 	\Pi_m = \{ \Psi\in C^{\infty}\ :\ 
 |\Psi(\sigma,t)|\leqslant P_m(\sigma t^{-2},
t^{-3}) 
 \ \,(\sigma,t)\in\mathbb{R}^1\times(0,T) 
 \} 
 $$ 
 ($P_m$ is a homogeneous polynomial of
degree~$m$). 
 
 Now, let us consider $\Psi^-_b(x,t,\varepsilon)$. 
 For $t>\varepsilon^{1/4-\lambda}$ 
 	$$ 
 	|\Psi^-_b(x,t,\varepsilon)|\leqslant |y^+(x,t)| 
 	\exp\left[ F^-(0,x,t)/\varepsilon \right] 
 	= O \left( \exp[-\varepsilon^{-\lambda_0}]\right), 
 \quad \lambda_0>0. 
 	$$ 
 For $t<\varepsilon^{1/4-\lambda}$ 
 we represent it in the form sum: 
 	$$ 
 	\Psi^-_b(x,t,\varepsilon) = 
 \int\limits_{y^+(x,t)}^{-gt^{\gamma}}
e^{F^-/\varepsilon} dy 
 + \int\limits_{-gt^{\gamma}}^{0} e^{F^-/\varepsilon} dy, 
 	$$ 
 where $g>0$, $3/2<\gamma<3/(1-\alpha)-2$. 
The first integral is estimated
as follows: 
 	$$ 
 	\left| \int\limits_{y^+(x,t)}^{-gt^{\gamma}} 
 e^{F^-/\varepsilon} dy \right| 
 \leqslant |y^+(x,t)| \exp\left[ F^-(-gt^{\gamma},x,t)/\varepsilon\right] 
 = O\left(e^{-g_{1} t^{2+\gamma}/\varepsilon}\right) . 
 $$ 
 The inequality $\gamma<3/(1-\alpha)-2$ provides 
an exponential smallness of the remainder
in the domain  
$$ 
 	\Omega_{\alpha}=\{ |\sigma| < t^2\varepsilon^{4\alpha-1}, 
 \ \varepsilon^{1-\alpha}<t^3<{\rm const},\ 0 < \alpha < 1 \}.
 $$ 
 From the inequality $\gamma>3/2$ we conclude that
for $-gt^{\gamma}\leqslant y \leqslant 0$ 
 and $t<\varepsilon^{1/4-\lambda}$ 
 	$$ 
 \delta(y,t) \equiv 
 	\frac{1}{\varepsilon}\left[ 
 	\frac{ty^2}{4(1+t)} + \frac{ay^3}{6} \right] 
 = O\left( \varepsilon^{\lambda_1}\right), 
 \quad\mbox{where}\quad \lambda_1>0. 
 	$$ 
 This allows one to expand $e^{F^-/\varepsilon}$
in the second integral into 
 a Taylor series in the small parameter $\delta(y,t)$:
 	$$ 
 	\Psi^-_{b} = \exp\left\{ -\displaystyle\frac{\zeta^2}{4(1+t)}\right\} 
 \left[ \sum\limits_{m=1}^{N-1}\varepsilon^{m} 
 \sum\limits_{l=0}^{[(m-1)/2]} \!\!\!
a_{m,l} 
 \frac{t^{m-1-2l} (1+t)^{m+l}}{x^{2m-1-l}} 
 	+ O\left(t \varepsilon^{\alpha N} \right) \right], 
  $$ 
 where, in particular, $a_{1,0} = 2$, $a_{2,0} = 4$. 
 Let us pass to variable $\sigma$: 
 	$$ 
 \Psi^-_{b} = 
 	\exp\left\{ -\displaystyle\frac{\zeta^2}{4(1+t)}\right\} 
 \left[ \frac{2\varepsilon(1+t)}{x} 
 + \sum\limits_{m=2}^{N-1}\varepsilon^{m} 
 \Psi^-_{b,m}(\sigma,t) 
 + O(\varepsilon^{\alpha N})\right], 
\qquad \Psi^-_{b,m}\in\Pi_m.
 $$ 

 Thus, we obtain
 	$$ 
 	\Psi^-(x,t,\varepsilon) = 
 \sqrt{2\pi\varepsilon} 
 \exp\left[ \frac{F^-(y^-(x,t),x,t)}{\varepsilon}\right] 
 \frac{\sqrt{6(1+t)}}{r(t)} + 
 $$ 
 	\begin{equation}\label{psim} 
 + \frac{2\varepsilon(1+t)}{x} 
 \exp\left\{ -\displaystyle\frac{\zeta^2}{4(1+t)}\right\} 
 + \exp\left\{ -\displaystyle\frac{\zeta^2}{4(1+t)}\right\} 
 O\left(t \varepsilon^{2\alpha} \right) . 
 \end{equation} 
 
The  function $\Psi^0$ can be represented
 in the form of the sum: 
 $\Psi^0 = \Psi^0_s + \Psi^0_b$, where 	$$ 
 \Psi^0_s(x,t,\varepsilon) 
 = \int\limits_{-\infty}^{y^{+}(x,t)}(y+ay^2)e^{F^-/\varepsilon} dy, 
 \qquad 
 \Psi^0_b(x,t,\varepsilon) 
 = \int\limits_{y^+(x,t)}^{0}(y+ay^2)e^{F^-/\varepsilon} dy. 
 $$ 
Functions $\Psi^0_s$ and $\Psi^0_b$ 
 are studied in exactly the same way as $\Psi^-_s$
and $\Psi^-_b$, respectively. 
 Thus, 	$$ 
 	\Psi^0_b(x,t,\varepsilon) = 
 \exp\left\{-\frac{\zeta}{4(1+t)} \right\} 
 \left[ 
 \sum\limits_{m=2}^{N-1}\varepsilon^{m} 
 \Psi^0_{b,m}(\sigma,t) 
 + O(\varepsilon^{\alpha N}) 
 \right], 
\qquad  \Psi^0_{b,m}\in\Pi_m,
 $$ 
 	$$ 
 	\Psi^0_s(x,t,\varepsilon) = - \sqrt{2\pi\varepsilon} 
 	\exp\left[ \frac{F^-(y^-(x,t),x,t)}{\varepsilon}\right] 
 \sum\limits_{j=0}^{\infty}\varepsilon^{j} \Psi^0_j(x,t), 
 	$$ 
 where 	$$ 
 \Psi^0_0(x,t) = \frac{t+R(x,t)+2ax(1+t)}{2a(1+t)^2} H(x,t), 
 	$$ 
 $$ 
 	\Psi^0_j(x,t) = q [H(x,t)]^{6j-3} + 
 q \frac{R(x,t)-1}{(1+t)} [H(x,t)]^{6j-1} 
 + q\frac{t+R(x,t)+2ax(1+t)}{(1+t)^2} [H(x,t)]^{6j+1}. 
 $$ 
 Passing to variable $\sigma$, we obtain 	\begin{equation}\label{psi0} 
 	\Psi^0_s(x,t,\varepsilon) = 
 - \sqrt{2\pi\varepsilon} 
 	\exp\left[ \frac{F^-(y^-(x,t),x,t)}{\varepsilon}\right] 
 \frac{\sqrt{6(1+t)}}{r(t)} 
 	\left\{ 2\mu(t) 
 	+ \sum\limits_{m=1}^{N-1}\varepsilon^{m} 
 \Psi^0_{s,m}(\sigma,t) 
 + O(\varepsilon^{\alpha N})\right\}, 
 	\end{equation} 
 where $\Psi^0_{s,m}\in\Pi_m$.

 Using expression~(\ref{psip}), (\ref{psim}),
and~(\ref{psi0}), 
 we arrive at the following result~\cite{vm4}. 

\vskip 3mm 
 \noindent 
 \textbf{Theorem~1.} 
 {\it In the domain 	
$$ 
 	\Omega_{\alpha}=\{ |\sigma| < t^2\varepsilon^{4\alpha-1}, 
 \ \varepsilon^{1-\alpha}<t^3<{\rm const},\ 0 < \alpha < 1 \}
 $$ 
 for the solution of problem $(\ref{eqw})$--$(\ref{ivp})$,
 there holds the asymptotic formula 
	\begin{equation}\label{hex} 
 	u(x,t,\varepsilon) = \sum\limits_{p=0}^{N-1} 
 \varepsilon^{p/2} h_p(\sigma,\zeta,t) 
 + O(\varepsilon^{\bar N}), 
 \end{equation} 
 where ${\bar N}\to\infty$ as $N\to \infty$. 
 The leading terms of this expansion is
 $$ 
 h_0(\sigma,\zeta,t) 
 = \frac{2\,\mu(t)}{	1 + K(t) \exp(\mu(t)\sigma) 
 \left[ 1 + \displaystyle\frac{1}{2\sqrt{\pi}} 
 \int\limits_{-\infty}^{-\zeta/(2\sqrt{1+t})} 
 \frac{e^{-z^2}}{z^2}dz 
 \right] }, 
 	$$ 
 	$$ 
 	\quad\mbox{where}\quad 
 	K(t) = \frac{r(t)}{\sqrt{3}} = 
 \sqrt{\frac{t}{2}} \left( 1 + \frac{3t}{32} + O(t^2) \right).
 	$$ 
 } 

Expansion  $(\ref{hex})$ shows the presence
of an additional scale in the form
of the stretched variable
$\zeta = x / \sqrt{\varepsilon}$.

 \bigskip 
\setcounter{equation}{0}
 \section{Large initial gradient} 
 
 In this section, we consider the Cauchy problem
for a more general quasilinear parabolic equation~\cite{sp,2ps,zvm,TP}: 
\begin{equation}\label{peq} 
 \frac{\partial u}{\partial t} + 
 \frac{\partial \varphi(u)}{\partial x} = 
 \varepsilon \frac{\partial^2 u}{\partial x^2},
 \quad t\geqslant 0,  \quad \varepsilon>0, 
 \end{equation}
\begin{equation}\label{iic}
 u(x,0,\varepsilon,\rho) = \nu ( {x}{\rho}^{-1}),
 \quad x\in\mathbb{R},  \quad \rho>0. 
 \end{equation}
 We assume that the function $\varphi$ is infinitely differentiable
 and its second derivative is strictly positive. 
 The initial function~$\nu$ is bounded and smooth. 
The interest to the problem under consideration is explained
by applications to studying processes
of shock waves formation~\cite{ib, FM, MO}.

The asymptotics of the solution
 as $\varepsilon\to 0$ and $\mu=\rho/\varepsilon\to 0$ 
 in the leading approximation has the form
$$ 
 {u}(x,t,\varepsilon,\rho)= 
 h_0\left(  \frac{x}{\rho}, 
 \frac{\varepsilon t}{\rho^2}\right)
- R_{0,0,0}\left( 
 \frac{x}{2\sqrt{\varepsilon t}} 
 \right)+ 
 \Gamma\left( 
 \frac{x}{\varepsilon}, 
 \frac{t}{\varepsilon} 
 \right) 
 +O\left(\mu^{1/2}\ln\mu\right), 
 $$ 
 where $h_0$, $R_{0,0,0}$ and $\Gamma$ are known functions.
Here, a multiscale character of asymptotics
arises from the beginning because of another
small parameter $\rho$.
 
 The behavior of the solution 
of problem $(\ref{peq})$--$(\ref{iic})$ is mainly 
determined by the solution of the limit problem 
\begin{equation}\label{lcp} 
 \frac{\partial u}{\partial t} + 
 \frac{\partial \varphi(u)}{\partial
x} =0, 
 \qquad 
 u(x,0) = 
 \begin{cases} 
 \nu^-_0, & x<0, \\ 
 \nu^+_0, & x\geqslant 0. 
 \end{cases} 
 \end{equation} 
 For $\nu^-_0>\nu^+_0$ using the method 
of characteristics,
 we find its generalized solution $$ 
 u_{0,0}(x,t) = 
 \begin{cases} 
 \nu^-_0, & x<ct, \\ 
 \nu^+_0, & x>ct, 
 \end{cases} 
 \qquad 
 {c}=  \frac{\varphi(\nu_0^+)-\varphi(\nu_0^-)}{\nu_0^+ -\nu_0^-}. 
 $$ 
This solution is discontinuous on the line of the shock wave $x=ct$.

In paper~\cite{TP}, 
for problem $(\ref{peq})$--$(\ref{iic})$
 outer expansions
\begin{equation}\label{outp} 
 U^+(x,t,\varepsilon,\rho) = \nu^+_0 + 
 \sum\limits_{m = 1}^{\infty}\sum\limits_{n=0}^{m-1} 
 \rho^{m-n}\varepsilon^{n} u^+_{m,n}(x,t),
 \end{equation} 
\begin{equation}\label{outm} 
 U^-(x,t,\varepsilon,\rho) = \nu^-_0 
 + \sum\limits_{m = 1}^{\infty}\sum\limits_{n=0}^{m-1} 
 \rho^{m-n}\varepsilon^{n} u^-_{m,n}(x,t)
 \end{equation} 
are constructed in domains
 $$ 
 \Omega^+_0 = \{ (x,t)\,: \, x>ct+\varepsilon^{1-\delta_0}, 
 \ 0<\delta_0<1\}
 $$ 
and
 $$ 
 \Omega^-_0 = \{ (x,t)\,: \, x<ct-\varepsilon^{1-\delta_0}\},
 $$ 
respectively, where
$$
 u^{\pm}_{m,n}(x,t) = 
 \sum\limits_{s=n}^{m-1} 
 \frac{ \alpha^{\pm}_{m,n,s}\, t^s}{[x-\varphi'(\nu^{\pm}_0)t]^{m+s}}, 
$$
 $\alpha^{\pm}_{m,n,s}$ are constants. 
 
Now let us construct an asymptotic solution
in a neighborhood of the line of discontinuity. 
First, rewriting outer expansions~(\ref{outp})
and~(\ref{outm})  in terms of the inner variable 
$$
 \sigma= \frac{x-ct}{\varepsilon},
$$
we obtain
 \begin{equation}\label{oer} 
 U^\pm(ct+\varepsilon\sigma,t,\varepsilon,\rho) = \nu^\pm_0 + 
 \sum\limits_{n=1}^{\infty} 
 \sum\limits_{m=1}^{[n/2]} 
 \mu^{m} \varepsilon^{n-m} 
 t^{m-n} P^\pm_{n-2m}(\sigma), 
 \end{equation} 
 where $P^\pm_{n-2m}(\sigma)$ are polynomials of
degree $n-2m$. 
 Taking into account 
the structure of series~(\ref{oer})
and the structure of the inner expansion
$$
 H=\sum\limits_{n=0}^{\infty} 
 \mu^{n} h_n(x/\rho,\omega), 
\qquad 
 \omega = t\varepsilon /\rho^2, $$
$$
 h_n =  \omega^{n/2} \sum\limits_{m=0}^{\infty} \omega^{-m/2} 
 \sum\limits_{l=0}^{m} (\ln\omega)^l 
 R_{n,m,l}\left(\frac{x}{2\rho\sqrt{\omega}}\right),
\quad |\sigma|+\omega\to \infty,
$$
we will construct an asymptotic solution of
equation~(\ref{peq}) 
 in the domain $$ 
 \Omega_3 = \{ (x,t)\,: \, 
 |x-ct|< \varepsilon^{1-\delta_3}, 
 \ t>\varepsilon^{1-\gamma_3}, 
 \ 0<\gamma_3,  \ \delta_0< \delta_3 <1\} 
 $$ 
 in the form of the  series 
$$ 
 V(\sigma,t,\mu,\varepsilon)= 
 v_0(\sigma) + 
 \sum\limits_{n=1}^{\infty} 
 \sum\limits_{m=1}^{n} 
 \mu^{m} \varepsilon^{n-m} 
 \sum\limits_{0\leqslant p+q\leqslant
n} 
 (\ln\mu)^{p} (\ln\varepsilon)^{q} 
 v_{m,n-m,p,q}(\sigma,t),
 $$ 
where
\begin{equation}\label{vzero} 
 v_0(\sigma)= \Lambda\left(\sigma+\varkappa\right)
 \end{equation} 
is the leading term,
the function $\Lambda$ is defined by the
formula $$ 
 \int\limits_{(\nu_0^+ + \nu_0^-)/2}^{\Lambda(\sigma)} 
 \frac{dv}{\varphi(v)-cv-b}=\sigma, 
 $$ 
 $$ 
 {c}= 
 \frac{\varphi(\nu_0^+)-\varphi(\nu_0^-)}{\nu_0^+ -\nu_0^-}, 
 \qquad 
 {b}= 
 \frac{\nu_0^+\varphi(\nu_0^-)-\nu_0^-\varphi(\nu_0^+)}{\nu_0^+ -\nu_0^-}, 
 $$ 
 $\varkappa$ is   constant,
which should be determined by the matching procedure.

In a standard way we arrive at the system of equations 
\begin{equation}\label{ev} 
 \frac{\partial^2 v_0 }{\partial \sigma^2} 
 +c\frac{\partial v_0 }{\partial \sigma}- \frac{\partial\varphi(v_0) }{\partial \sigma}=0, 
 \end{equation} 
 \begin{equation}\label{10pq} 
 L_3 v_{1,0,p,q} = 0, 
 \end{equation} 
 \begin{equation}\label{mnpq} 
 L_3 v_{m,n-m,p,q} = \frac{\partial
v_{m,n-1-m,p,q}}{\partial t} 
 +\frac{\partial Q_{m,n,p,q}}{\partial \sigma}, 
 \end{equation} 
 where 
\begin{equation}\label{l3} 
 L_3v\equiv \frac{\partial^2 v }{\partial \sigma^2} 
 +c\frac{\partial v }{\partial \sigma}- \frac{\partial\varphi'(v_0) v }{\partial \sigma}, 
 \end{equation} 
 \begin{equation}\label{vrh} 
 Q_{m,n,p,q} = \sum\limits_{r=2}^{n} 
 \frac{\varphi^{(r)}(v_0)}{r!} 
 \sum\limits_{\mathfrak{S}} 
 \prod\limits_{k=1}^{r} 
 v_{m_k,n_k-m_k,p_k,q_k}, 
 \end{equation} 
 $$ 
 \mathfrak{S} = 
 \{ (m_k,n_k,p_k,q_k):\, 
 \sum\limits_{k=1}^{r} m_k=m, 
 \sum\limits_{k=1}^{r} n_k=n, 
 \sum\limits_{k=1}^{r} p_k=p, 
 \sum\limits_{k=1}^{r} q_k=q\}.
 $$ 
 
We should find solutions of 
the obtained system satisfying
the conditions 
\begin{equation}\label{vms} 
 v_{m,s,p,q}(\sigma,t)= V^\pm_{m,s,p,q}(\sigma,t) 
 +O(\sigma^{-\infty}), 
 \qquad \sigma\to\pm\infty, 
 \end{equation} 
 where 
\begin{equation}\label{vd} 
 V^\pm_{m,s,p,q}(\sigma,t) 
 = 
 \begin{cases} 
 t^{-s} P^\pm_{s-m}(\sigma), & s\geqslant m,\
p=q=0, \\ 
 0, & \text{otherwise}, 
 \end{cases} 
 \end{equation} 
the symbol $O(\sigma^{-\infty})$ denotes a function of
order $O(\sigma^{-N})$ for any $N>0$.

 Denote by $\mathfrak{C}$ 
the class of $C^{\infty}$-smooth functions $v(\sigma,t)$ 
for $\sigma\in\mathbb{R}$, $T_1\leqslant t \leqslant T_2$. 
By $\mathfrak{M}^+$ we denote the set of functions
from $\mathfrak{C}$ such that 
there hold the inequalities
 $$ 
 \left| \frac{\partial^{i+j} v(\sigma,t)}{\partial \sigma^{i}\partial t^j} 
 \right| 
 \leqslant M_{i,j} \exp(-\gamma\sigma), 
 \qquad 
 \forall i,j. 
 $$ 
 By $\mathfrak{M}^-$ we denote an analogous set of
function, for which there hold the
inequalities
 $$ 
 \left|\frac{\partial^{i+j} v(\sigma,t)}{\partial \sigma^{i}\partial t^j} 
 \right| 
 \leqslant M_{i,j} \exp(\gamma\sigma), 
 \qquad 
 \forall i,j. 
 $$ 
 
 Further, we need the following statement. 

\vskip 3mm
\noindent
\textbf{Lemma~1.}
{\it  Let $$ 
 P^{-}\in\mathfrak{C}, 
 \qquad 
 P^{+}\in\mathfrak{C}, 
 \qquad 
 F\in\mathfrak{C}, 
 $$ 
 $$ 
 {F}-L_3 P^{-}\in\mathfrak{M}^{-}, 
 \qquad 
 {F}-L_3 P^{+}\in\mathfrak{M}^{+}. 
 $$ 
 Then for the existence of a 
solution of the problem $$ 
 L_3 v = F, 
 \qquad 
 {v}-P^{\pm}\in\mathfrak{M}^{\pm} 
 $$ 
it is necessary and sufficient
 the fulfillment of the condition
\begin{equation}\label{exco} 
 \begin{split} 
 &\left\{\frac{\partial}{\partial \sigma} 
 (P^+ - P^-) + [c-\varphi(v_0)](P^+ - P^-) 
 \right\}_{\sigma=0}= \\ 
 &=\int\limits_{-\infty}^{0} 
 [F(\sigma,t)-L_3 P^-(\sigma,t)]d\sigma+ 
 \int\limits_{0}^{\infty} 
 [F(\sigma,t)-L_3 P^+(\sigma,t)]d\sigma. 
 \end{split} 
 \end{equation} 
} 
The proof of this lemma
can be found in~\cite[chapter VI]{ib}. 

\vskip 3mm 
\noindent
\textbf{Theorem~2.}
{\it
 For $\sigma\in\mathbb{R}$ and $t>0$ there exist
solutions of equations $(\ref{ev})$--$(\ref{mnpq})$
 such that $\ v_0-\nu_0^\pm\in\mathfrak{M}^{\pm}$
and ${v}_{m,n-m,p,q}-V_{m,n-m,p,q}^{\pm}\in\mathfrak{M}^{\pm}$, 
 where $V_{m,n-m,p,q}^{\pm}$ are functions $(\ref{vd})$. 
Under the condition  that all ${v}_{m',n'-m',p',q'}$ for $n'<n$
are already determined,
each function ${v}_{m,n-m,p,q}$ is determined
uniquely up to a term ${\varkappa}_{m,n-m,p,q}v'_0(\sigma),$ 
where ${\varkappa}_{m,n-m,p,q}$ is an arbitrary constant. 
} 

\noindent 
\textbf{Proof.}
Formula~(\ref{vzero}) gives
the function $v_0(\sigma)$
 satisfying equation~(\ref{ev}) 
 and condition $v_0-\nu_0^\pm\in\mathfrak{M}^{\pm}$.
Thus, problem~(\ref{10pq}), (\ref{vms}) has
the solution $v_{1,0,p,q}(\sigma,t) = {\varkappa}_{1,0,p,q}(t)v'_0(\sigma)$. 
Proceeding by induction,
suppose that solutions of problems~(\ref{mnpq}), (\ref{vms}) 
 for $n<k$ are constructed so that for $n=k$
the problem has a solution $$ 
 v^{*}_{m,k-m,p,q}(\sigma,t) + {\varkappa}_{m,k-m,p,q}(t)v'_0(\sigma). 
 $$ 
 Let us find a function ${\varkappa}_{m,k-m,p,q}(t)$
for which problem~(\ref{mnpq}), 
 (\ref{vms}) has a solution for $n=k+1$. 
Apply Lemma~1,  whose conditions are fulfilled for $$ 
 F(\sigma,t) = 
 \frac{\partial v_{m,k-m,p,q}}{\partial
t} 
 +\frac{\partial Q_{m,k+1-m,p,q}}{\partial \sigma}, 
 \quad 
 P^{\pm}(\sigma,t) = V^{\pm}_{m,k+1-m,p,q}(\sigma,t),
 $$ 
to the function $v_{m,k+1-m,p,q}(\sigma,t)$.
 According to Lemma~1, for the solvability of
problems~(\ref{mnpq}), (\ref{vms}) for $n=k+1$ 
it is necessary and sufficient the fulfillment of
conditions~(\ref{exco}). 
 Substituting functions $v_{m,n-m,p,q}(\sigma,t)$ for $n<k$
and $$ 
 v_{m,k-m,p,q}(\sigma,t) = v^{*}_{m,k-m,p,q}(\sigma,t) + 
 {\varkappa}_{m,k-m,p,q}(t)v'_0(\sigma)
 $$ 
into this equality,
 we see that the function ${\varkappa}_{m,k-m,p,q}(t)$
enters in the form $$ 
 \int\limits_{-\infty}^{0} 
 \frac{\partial({\varkappa}_{m,k-m,p,q}(t)v'_0(\sigma))}{\partial t} 
 d\sigma+ 
 \int\limits_{0}^{\infty} 
 \frac{\partial({\varkappa}_{m,k-m,p,q}(t)v'_0(\sigma))}{\partial t} 
 d\sigma = (\nu_0^+ -\nu_0^-) 
 \frac{d{\varkappa}_{m,k-m,p,q}(t)}{dt}. 
 $$ 
 Thus, condition~(\ref{exco}) has the
form 
$$
{\varkappa'}_{m,k-m,p,q}(t)=f(t),$$
 where $f(t)$ is a known function. 
 This implies that a solution $v_{m,k-m,p,q}(\sigma,t)$ is
determined  up to a term ${\varkappa}_{m,k-m,p,q}v'_0(\sigma)$
 and problem~(\ref{mnpq}), (\ref{vms}) is solvable
for $n=k+1$. 
 Theorem~2 is proved.

 \bigskip 
\setcounter{equation}{0}
\section{Dispersive compression wave} 

Dispersive compression waves are studied
in plasma, fluids, and optics~\cite{TBI, Lh, JWF,CFP}.
In particular, studying properties of solutions 
 of the Korteweg--de Vries equation 
 \begin{equation}\label{eq} 
 \frac{\partial u}{\partial t} + 
 u\frac{\partial u}{\partial x} + 
 \varepsilon \frac{\partial^3 u}{\partial x^3} = 0, 
 \quad t\geqslant 0, \quad \varepsilon>0, 
 \end{equation} 
 is of indisputable interest 
 for describing nonlinear wave phenomena. 
For steplike initial data, an asymptotic 
solution of the KdV equation (\ref{eq})
 can be found by the inverse spectral transform method~\cite{Kh}.
 Under certain restrictions on the initial function,
 the asymptotic behavior can be studied by the Whitham method,
 as in~\cite{gke,kke}.
The long-time asymptotic solution with step-like initial data 
is also analyzed in~\cite{AB1,AB2}.

 Open questions concerning the behavior of
solutions are still the subject of attention 
 for modern researches~\cite{gst}. 
 Some mathematical results about the solution of
the problem in various cases can be found
in~\cite{kf} and~\cite{fam}.

Here, we consider results of paper~\cite{tmf} for the Cauchy problem 
 with the initial condition 
 \begin{equation} \label{ic}\phantom{\frac{1}{1}} 
 u(x,0,\varepsilon,\rho) = \Lambda ( {x}{\rho}^{-1}), 
 \quad t=0, \quad x\in\mathbb{R}, \quad \rho>0, 
 \end{equation} 
introduced in the previous section.
We assume the fulfillment of the
condition 
 $$\mu=\displaystyle\frac{\rho}{\sqrt{\varepsilon}}\to 0.$$ 
 
 In paper~\cite{tmf}, an asymptotic  solution to problem $(\ref{eq})$--$(\ref{ic})$ 
 is constructed using renormalization
\cite{bk2,gv,teodor}. 
 Let us pass to the inner variables $$ 
 x = \sqrt{\varepsilon}\,\eta, 
 \qquad 
 t = \sqrt{\varepsilon}\,\theta, 
 $$ 
 since this allows one to take into account all
terms in equation (\ref{eq}). 
 As a first approximation, we take the solution
of 
 the equation 
 \begin{equation}\label{geq} 
 \frac{\partial Z}{\partial \theta} + 
 Z\frac{\partial Z}{\partial \eta} + 
 \frac{\partial^3 Z}{\partial \eta^3} = 0, 
 \end{equation} 
 with the initial condition 
 \begin{equation}\label{gic} 
 Z(\eta,0) = 
 \begin{cases} 
 \Lambda^-_0, & \eta<0, \\ 
 \Lambda^+_0, & \eta > 0, 
 \end{cases} 
 \end{equation} 
 where $\Lambda^\pm_0=\lim\limits_{s\to\pm\infty}\Lambda(s)$. 
 As a model of collisionless shock waves, 
 problem~(\ref{geq})--(\ref{gic}) 
 was studied by A.V. Gurevich and L.P. Pitaevskii 
 in~\cite{gp2}. 
 
 Let us construct the expansion of the solution in
the 
 following form: 
 \begin{equation}\label{rge} 
 u(x,t,\varepsilon,\rho) = 
 Z(\eta,\theta) + \mu W(\eta,\theta,\mu) + O(\mu^{\alpha}), 
 \qquad \alpha>0, 
 \end{equation} 
 where the addend $\mu W(\eta,\theta,\mu)$ must
eliminate 
 the singularity of $Z$ at the initial moment of
time. 
 Then the function~$W$ satisfies the
linear 
 equation 
 \begin{equation}\label{wle} 
 \frac{\partial W}{\partial \theta} + 
 \frac{\partial (Z W)}{\partial \eta} + 
 \frac{\partial^3 W}{\partial \eta^3} = 0. 
 \end{equation} 
 Differentiating equation~(\ref{geq}) 
 with respect to $\eta$, we find that 
 the expression $$ 
 G(\eta,\theta)= 
 \frac{1}{\Lambda_0^+ - \Lambda_0^-} 
 \frac{\partial Z(\eta,\theta)} 
 {\partial \eta} 
 $$ 
 satisfies equation~(\ref{wle}). 
 Moreover, $G$ is the Green function, 
 because $$ 
 \lim\limits_{\theta\to+0} 
 \int\limits_{-\infty}^{\infty} 
 G(\eta,\theta)f(\eta)\, d\eta = 
 - \frac{1}{\Lambda_0^+ - \Lambda_0^-} 
 \int\limits_{-\infty}^{\infty} 
 Z(\eta,0) f'(\eta)\, d\eta = f(0) 
 $$ 
 for any smooth function~$f$ with compact
support, 
 thus $G(\eta,0)=\delta (\eta)$.

 Let us choose the solution~$W$ in the
form 
 the convolution 
 with the Green function~$G$ 
 so that the asymptotic approximation would
satisfy  the initial condition~(\ref{ic}). 
 As a result, expansion~(\ref{rge})
becomes 
$$ 
 u(x,t,\varepsilon,\rho) = U_0(x,t,\varepsilon,\rho) 
 + O(\mu^{\alpha}), 
 $$ 
 where 
 $$ 
 U_0(x,t,\varepsilon,\rho) = Z(\eta,\theta) + 
 \frac{\mu}{\Lambda^+_0 - \Lambda^-_0} 
 \int\limits_{-\infty}^{\infty} 
 \frac{\partial Z(\eta - \mu s,\theta)}{\partial \eta} 
 \left[ \Lambda(s) - Z(s,0)\right]\, ds. 
 $$ 
 Integrating by parts, we obtain the 
 asymptotic approximation 
$$
 u(x,t,\varepsilon,\rho) \approx 
 U_0(x,t,\varepsilon,\rho) = 
 \frac{1}{\Lambda^+_0 - \Lambda^-_0} 
 \int\limits_{-\infty}^{\infty} 
 Z\left( \frac{x-\rho s}{\sqrt{\varepsilon}}, \frac{t}{\sqrt{\varepsilon}}\right) 
 \Lambda'(s)\, ds. 
$$
 
 Constructing complete asymptotic expansions 
 of the solution near the singular point 
 by the standard matching method 
 may be connected with serious difficulties. 
 In fact, it is necessary to solve 
 the scattering problem for a recurrence
system 
 of partial differential equations 
 with variable coefficients~\cite{ib}. 
 In addition, the investigation of the shock
wave 
 generated by gradient catastrophe shows
that 
 the asymptotics of the solution in a
neighborhood 
 of a singular point may have a multiscale
structure~\cite{vm4} as we have seen in Section~1.
 
 The renormalization approach allows one
to 
 construct a uniformly suitable asymptotics 
 in the whole domain of independent
variables 
 avoiding difficulties arising 
 from the matching procedure. 

In particular,
 for $\Lambda^+_0=0$ and $\Lambda^-_0=a>0$
the following formula was obtained in paper~\cite{tmf}:
$$
u(x,t,\varepsilon,\rho) \approx
2\Lambda\left( \frac{x+at}{\rho}\right)
-\Lambda\left( \frac{x-2at/3}{\rho}\right)-
$$
$$
-\frac{at}{\rho}\int\limits_{-1}^{2/3}
\Lambda'\left( \frac{x- aty}{\rho}\right)
\left[2\, \mathrm{dn}^2
\left(\frac{a^{3/2} t\,\omega(y)}{\sqrt{\varepsilon}},
\sigma(y)\right) 
+\sigma^2(y)
\right] dy,
$$
where  $\mathrm{dn}(v,\sigma)$ is the elliptic Jacobi function
$$
\mathrm{dn}(u,m)= \sqrt{1-m\sin^2\varphi},
\qquad
u=\int\limits_{0}^{\varphi(u)}
\frac{dv}{\sqrt{1-m \sin^2 v}},
$$
$$
\omega(y)= \frac{1}{\sqrt{6}}
\left\{ y - \frac{1}{3}\left[ 1+\sigma^2(y)\right] \right\},
\qquad
1+\sigma^2-
\frac{2\sigma^2 (1-\sigma^2)K(\sigma)}
{E(\sigma)-(1-\sigma^2)K(\sigma)}=3y,
$$
$K(\sigma)$ and $E(\sigma)$ are
complete elliptic integrals of first and second kind:
$$
K(\sigma) = \int\limits_{0}^{\pi/2}
\frac{dv}{\sqrt{1-\sigma^2 \sin^2 v}}
\qquad
E(\sigma) = \int\limits_{0}^{\pi/2}
\sqrt{1-\sigma^2 \sin^2 v}\, dv.
$$


\begin{thebibliography}{99}

\bibitem{bu}
J. Burgers, \textit{A Mathematical Model Illustrating the Theory of Turbulence},
 Advances in Applied Mechanics, Academic
Press, New York, 1948.

\bibitem{Gbw}
 G.~B. Whitham, \textit{Linear and Non-Linear Waves}, Wiley-Interscience, 
New York, 1974. 

\bibitem{ib}
 A.M. Il'in,
\textit{Matching of Asymptotic Expansions of Solutions of Boundary Value Problems},
AMS, Providence, RI, 1992. 

\bibitem{vm4}
 S.V. Zakharov,
\textit{Comp. Math. Math. Phys.}, 44:3 (2004), 506-513. 

\bibitem{sp}
S.V. Zakharov,
arXiv:math-ph/1411.4395v1  (2014).

URL: http://arxiv.org/abs/1411.4395

\bibitem{BKdV}
S.V. Zakharov,
arXiv:math-ph/1501.02542v1 (2015).

URL: http://arxiv.org/abs/1501.02542.

\bibitem{DM1}
V. Danilov, D. Mitrovic,
arXiv:math.AP/0405325v1 (2004).

\bibitem{DSh}
V.G. Danilov, V.M. Shelkovich,
 Nonlinear Studies, 8 (2001), no. 1, 211-245.

\bibitem{2ps}
{S.V. Zakharov,}
{\it  Asymptotic Analysis},  \textbf{63} No.1-2. (2009), 49-54.

\bibitem{zvm}
 S.V. Zakharov,
\textit{Comp. Math. Math. Phys.}, 50:4 (2010), 665-672. 

\bibitem{TP}
S.V. Zakharov,
arXiv:math.AP/1504.04928v1 (2015).

URL: http://arxiv.org/abs/1504.04928.

\bibitem{FM}
L. Fan, A. Matsumura,
arXiv:math.AP/1307.3688v1 (2013).

\bibitem{MO}
R.W. Minich, D. Orlikowski, 
\textit{ The Symmetry and Geometry of Shock Formation},
 in ``New Models and Hydrocodes for Shock Wave Processes in
Condensed Matter'', May 24-28, 2010, Paris, France.

\bibitem{TBI}
 R. J. Taylor, D. R. Baker, H. Ikezi, 
\textit{Phys. Rev. Lett.} 24 (1970) 206-209.

\bibitem{Lh}
J. Lighthill,
\textit{Waves in Fluids}, Cambridge University Press, 1978.

\bibitem{JWF}
S. Jia, W. Wan, J. W. Fleischer,
\textit{ Phys. Rev. Lett.} 99 (2007) 223901.

\bibitem{CFP}
C. Conti, A. Fratalocchi, M. Peccianti, G. Ruocco, S. Trillo,
\textit{ Phys. Rev. Lett.} 102 (2009) 083902.

\bibitem{Kh}
E.Ya. Khruslov,
\textit{ Math. USSR-Sb.}, 28, 229-248 (1976).

\bibitem{gke}
A.V. Gurevich, A.L. Krylov, G.A. El',
\textit{JETP Lett.}, 54, 102-107 (1991).

\bibitem{kke}
A.L. Krylov, V.V. Khodorovskii, G.A. El',
\textit{JETP Lett.}, 56, 323-327 (1992).

\bibitem{AB1}
M.J. Ablowitz,
\textit{Nonlinear Dispersive Waves, Asymptotic Analysis and
Solitons}, Cambridge University Press, 2011.

\bibitem{AB2}
M.J. Ablowitz, D.E. Baldwin,
arXiv:1301.1032v1 [nlin.PS] (2013).

\bibitem{gst}
 R. Garifullin, B. Suleimanov, N. Tarkhanov,
\textit{Phys. Lett. A}, 374:13 (2010), 1420-1424. 

\bibitem{kf}
{T.~Kappeler},
\textit{J. Diff. Eq.}
63:3, (1986), 306-331.

\bibitem{fam}
A.V. Faminskii, 
\textit{Math. USSR-Sb.} 68:1 (1991), 31-59. 

\bibitem{tmf}
 S.V. Zakharov,
\textit{Theor. Math. Phys.}, 175:2 (2013), 592-595.

\bibitem{bk2}
 J. Bricmont, A. Kupiainen,
\textit{Renormalizing Partial Differential Equations},
 Lecture Notes in Physics, Springer-Verlag, 1994. 

\bibitem{gv}
 N. Goldenfeld, J. Veysey,
\textit{Rev. Mod. Phys.}, 79 (2007), 883-927. 

\bibitem{teodor}
 E.V. Teodorovich,
\textit{J. Appl. Math. Mech.}, 68:2 (2004), 299-326. 

\bibitem{gp2}
 A.V. Gurevich, L.P. Pitaevskii,
\textit{Sov. Phys.-JETP}, 38:2 (1974), 291-297. 

\end{thebibliography}
\end{document}